\documentclass{sig-alternate}
\usepackage{times}

\usepackage{pifont}
\usepackage{epsfig}
\usepackage{tabularx}
\usepackage{amssymb}
\usepackage{latexsym}
\usepackage{amsmath}
\usepackage{delarray}
\usepackage{moreverb}
\usepackage{xspace}

\usepackage{lscape}
\usepackage{changebar}
\usepackage{graphicx}
\usepackage{graphics}
\usepackage{mflogo}
\usepackage{xspace}
\usepackage{texnames}
\usepackage{rotating}
\usepackage{alltt}

\usepackage{amsmath}
\usepackage{epsfig}
\usepackage{booktabs,paralist}

\def\x{{\mathbf x}}

\newcommand{\Vc}[1]{{\bf #1}}

\newcommand{\mypar}[1]{{\bf #1.}}

\newcommand{\Singlefigure}[4]{{\begin{figure*} %
\centering %
\includegraphics[width=#1\linewidth]{#2}
\caption{#3}%
\label{#4}%
\end{figure*}}}

\graphicspath{{Figures/}}

\begin{document}

\title{Multiple-Campaign Ad-Targeting Deployment} 

\subtitle{Parallel Response Modeling, Calibration and Scoring
  \\ Without Personal User Information}

\numberofauthors{1} \author{ \alignauthor Paolo D'Alberto
  \\ \affaddr{NinthDecimal} \\ \email{\small pdalberto@ninthdecimal.com} }

\maketitle

\begin{abstract}
We present a vertical introduction to campaign optimization: we want
the ability to predict any user response to an ad campaign without any
users' profiles on an average and for each exposed ad, we want to
provide offline modeling and validation, and we want to deploy
dedicated hardware for on-line scoring. In practice, we present an
approach how to build a polytomous model (i.e., multi response model)
composed by several hundred binary models using generalized linear
models. The theory has been introduced twenty years ago and it has
been applied in different fields since then. Here, we show how we can
optimize hundreds of campaigns and how this large number of campaigns
may overcome a few characteristic caveats of single campaign
optimization. We discuss the problem and solution of training and
calibration at scale. We present statistical performance as {\em
  coverage}, {\em precision} and {\em recall} used in off line
classification. We present also a discussion about the potential
performance as throughput for on-line scoring: how many decisions can
be done per second streaming the bid auctions also by using dedicated
hardware such as Intel Phi.
\end{abstract}




\section{Introduction} 
\label{sec:introduction}

Free contents and targeted advertising are Janus-faces; that is, the
beginning and the end of a cycle. We start by watching original
contents and exposing our likes to the contents providers. In turn,
our preferences are used to provide audience targeting for
advertisers, which at the end of the day pay for the contents.

We would like that the advertising should be tailored to the audience
in such a way that the ads are relevant, engaging, and with high
precision in order to minimize the costs and to maximize the return to
the original investment.

Collecting personal history is a common approach used to understand
the audience and to find ways to reach them. Every time we log into
our mail account(s), social network(s), search(es) and for anything we
do afterwards, these events help describing who we are and what we
do. These profiles attract a large spectrum of interests as the like
of advertisers, recruiters, prospects, or at worst soliciting email.

Considering that a social network can reach billions of people,
collecting and classifying these volume of data is not an easy chore.
Nonetheless, the classification changes little in time. For example,
we like Italian Opera and we will, likely, for life; other likes may
be more flaky.

Now let us put the previous problem in the context of avail ads. For
example, there are ad-exchanges where an ad-company can go, bid, and
buy impressions (ad-space): by counting the largest exchanges, we
could bid up to 10 Billions impressions every day in US alone,
reaching 150 Million people, say, and serving thousands of
campaigns. If we deploy users' profiles, we need to have information
{\em only} about the 150 Millions. If we do not have such information,
we may have to make a new and a different decision for each
impression: the problem space becomes larger and thus the problem
harder (10 Billions).

In practice, not everyone wants to collect personal information and
not every one can or will be able to. One day, {\em cookies}
associated to browsers may be gone and thus the main mean to follow us
is to log in (Google, Facebook and Yahoo) or tracking the location of
our devices (Geo location). In this work, we do not collect any
personal information: we care about the collective and anonymous
response. We shall clarify what we collect in Section
\ref{sec:dimensions}.

The next question will be how can we tailor the response of hundreds
of campaigns? Campaigns are now being designed for an interaction with
the audience: for example, clicking to redeem a coupon, responding to
a trivia question, playing a game and others. These interactions set
apart these impressions: campaigns and advertisers want to respond to
the feedback in order to focus more to who really show interests.

Click Through Rate (CTR) is a measure of performance for a
campaign. By construction, CTR is a ratio defined as the number of
clicks over the total number of impressions delivered. Assume we reach
all members of our audience and the people really interested to the
campaign clicked to their satisfaction. If we have to redo the same
campaign, we would be parsimonious and reduce the number of
impressions maintaining the same interests, same number of clicks, and
thus larger CTR. 

\vspace{0.3cm}
\mypar{Problem} For example, given an impression $x$, with its
features from a pool of $n$ impressions, we would like to compute the
probability $P_{C_\ell}(x)$ of a click for the campaign $C_\ell$,
where we have $1\leq \ell \leq N$. In practice, we model this problem
by estimating each binomial distribution: {\footnotesize
\begin{alignat}{2}
\label{eq:distribution}
  P_{C_\ell}[Y=k] = \binom{n}{k}p_{C_\ell}^k(1-p_{C_\ell})^{n-k},
  \qquad k\in[0,n]
\end{alignat}
} We are after the parameter $p_{C_\ell}$, the average probability of
a click, and its connection with the features ${\bf x}$ of the
impression $x$; that is, $p_{C_\ell} = g({\bf x})$.

We organize the paper as follows.  In Section
\ref{sec:glm-introduction}, we introduce the basic applied statistic
method used to estimate the connection between any impression $x$ and
its probability as in Equation \ref{eq:distribution}.  In practice, we
dwell with a problem that has a small number of positive cases and a
very large number of negative ones. In Section
\ref{sec:positive-and-negative-set}, we present our approach to select
retrospective sampling, {\em ex post}, and in particular how to select
the negative cases (no real ones) and how to scale to provide
prospective estimate, {\em ex ante}, which can be used for bidding in
real time. We explain also our approach to calibrate the system and
choose thresholds to mimic a binary response. In Section
\ref{sec:dimensions}, we present how we explore the features space. In
Section \ref{sec:overall}, we put everything together and we present
measure of {\em quality} such a precision. Eventually, the computation
of $P_{C_\ell}(x)$ has to be efficient and performed at run time; in
Section \ref{sec:speed}, we provide throughput using state of the art
systems. We wrap up in Section \ref{sec:conclusions} and we tip our
hats to who helped us in Section \ref{sec:ack}.

\section{GLM,  Generalized Linear Models}
\label{sec:glm-introduction}
Suppose that we want to have a response $Y_i$ that can take only two
possible values: $Y_i=1$ and $Y_i=0$. For example, the former
represents the response of a click or success and the latter the
response to a non click or failure. We write
\begin{alignat}{2}
\label{eq:probability}
  P[Y_i=0] = 1-\pi_i, \qquad P[Y_i=1] = \pi_i
\end{alignat}
We describe this as $Y_i \sim B(1,\pi_i)$. This is a single trial. If
we have $m$ independent trials, and there is a common probability
$\pi$ of success and probability $1-\pi$ of failure, we say that $Y
\sim B(m,\pi)$ and $P[Y=k] = \binom{m}{k}\pi^k(1-\pi)^{n-k}$, for any
integer $1<k<m$. The probability distribution is counting the number
of $k$ possible successes over a pool of $m$ trials. The Binomial
distribution is one of the oldest to be studied and it was derived by
Jacob Bernoulli \cite{Bernoulli1713}. Our notations and references are
from \cite{JohnsonKK1993}, Chapther 2.

In general, we cannot assure homogeneity and we should consider a
process such as $Y = \sum_{i=1}^m Y_i$ where $Y_j \sim B(1,\pi_j)$,
that is the summation of non homogeneous Binomial. Also, the values of
$\pi_1, \dots, \pi_m$ are often unknown and we will eventually compute
an average $Y_i \sim B(1,\overline{\pi})$ and $Y \sim
B(m,\overline{\pi})$ where $\overline{\pi} =
\frac{1}{m}\sum_{i=1}^m\pi_i$.

It should be clear now that $Y$ is a response to an event. The event
is represented by a vector of explanatory variables
$\Vc{x}=(x_1,\dots,x_p)$. The principal objective of a statistical
analysis is to investigate the relationship between the response
probability $\pi$ and the explanatory $\Vc{x}$, that is to find
$g(\pi) \sim \Vc{x}$.

What follows, especially the notations and the meaning behind the
notation is from \cite{McCullaghN1989} Chapter 2 and 4.  Linear models play
an important role in applied and theoretical work. We suppose there
is a linear dependence
\begin{equation}
\label{eq:linear}
g(\pi_i) = \eta_i=\sum_{j=1}^p x_{i,j}\beta_j
\end{equation}  
for to be computed $\beta_1,\dots,\beta_p$. The function $g()$ is a
transformation that makes possible to map the range $[0,1]$, which is
the range of probability, to a more appropriate space
$(-\infty,+\infty)$, which is appropriate for a linear function. 

In this work we use the logistic function 
\begin{equation}
\label{eq:logit}
\log_e\Big(\frac{\pi_i}{1-\pi_i}\Big) =\beta_0 + \sum_{j=1}^p x_{i,j}\beta_j
\end{equation}
where the fraction $\frac{\pi_i}{1-\pi_i}$ has range in the interval
$[0,+\infty)$ and it also know as {\em odd ratio} (e.g., used for
  hypothesis testing and sequential analysis \cite{Wald1947}). In
  Equation \ref{eq:logit}, we single out the parameter $\beta_0$,
  which is the {\em intercept}.  As a note, the equation
  \ref{eq:logit} is for every trial and our ability of estimating
  $\pi_i$ however the unknown $\beta$s are determined for all. There
  is also another and very important reason to use the logistic
  regression as we shall explain in Section \ref{sec:post-vs-ante}.

\subsection{Computing $\Vc{\beta}$: Maximum likelihood.}

The responses $y_1,\dots,y_n=\Vc{y}$ are observed from the independent
binomial variables $Y_1,\dots,Y_n=\Vc{Y}$ such that $Y_i \sim
B(m_i,\pi_i)$ and we could use the expression $\Vc{Y} \sim
B(\Vc{m},\Vc{\pi})$.

The {\em log likelihood} may be written as:
\begin{equation}
\label{eq:loglikelihood}
l(\Vc{\pi},\Vc{y}) = \sum_{i=1}[y_i\log(\frac{\pi_i}{1-\pi_i})+m_i\log(1-\pi_i)]
\end{equation}
where we omit the term $\sum\log\binom{m_i}{y_i}$, which is a constant
independent of $\Vc{\pi}$. If we substitute the linear logistic model
in Equation \ref{eq:logit} we have 
\begin{equation}
\label{eq:logbeta}
l(\Vc{\beta},\Vc{y}) = \sum_i\sum_{j} y_i x_{i,j} \beta_j - \sum_i m_i \log(1+e^{\sum_j x_{i,j}\beta_j}),
\end{equation}
or in matrix form
\begin{equation}
\label{eq:logbetamatrix}
l(\Vc{\beta},\Vc{y}) = \Vc{y}^t \Vc{X}\Vc{\beta} - \sum_i m_i \log(1+e^{\Vc{X}_i^t\Vc{\beta}})
\end{equation}
 
We identify with $\hat{\Vc{\beta}}$ the value of $\Vc{\beta}$ that
maximizes Equation \ref{eq:logbeta}. The appealing of this formulation
is that log likelihood depends on $\Vc{y}$ only through the linear
combinations $\Vc{y}^t\Vc{X}$ and $\Vc{y}^t\Vc{X} =
E[\Vc{Y}^t\Vc{X};\hat{\Vc{\beta}}]$. The details of the computation of
$\hat{\Vc{\beta}}$ are available \cite{McCullaghN1989} Section
4.4.2. It is an iterative solver where at every iteration involves a
matrix factorization. The matrix factorization may be prohibitive as
the matrix $\Vc{X}$ will become larger.

The iterative approach uses weights to give more importance to
specific dimensions. The weight may adapt at each iterations and it
may require a computation of a different $\Vc{Q}\Vc{R}$ matrix
factorization. There are suggestions where the factorization can be
done once and being reused (circumventing one of the most expensive
step) because the weights will change the spectrum of
$\Vc{R}\sim\sqrt{\Vc{w}^t}\Vc{R}$ but will not change the spectrum of
$\Vc{Q}$, which must be a unitary matrix (e.g. \cite{golubVL1996}
Chapter 5).

\subsection{Ex Post vs. Ex Ante: different Intercept.} 
\label{sec:post-vs-ante}

In \cite{McCullaghN1989} Section 4.3.3, there is a full explanation
but this property should strike a chord to anyone using models for
predictions. In general, we use the past responses (i.e., ex post) to
compute $\Vc{\hat{\beta}}$: We consider the past clicks and non clicks
for the impressions we have delivered. If we are in the process of
completion we have not seen all clicks and, most importantly, we have
not seen all impressions yet. As matter of fact, the pool of available
impressions is even larger than the one we are going to deliver (i.e.,
ex ante).

The ex-post building model approach is practical. The ex ante is
not. If we could build both they will differ only by their intercept
$\beta_0$. This is true because of {\em logit} canonical link in
Equation \ref{eq:logit}. The difference can be computed if we have at
least an estimate of the difference in scale of the ex-post (training)
and ex-ante (total) sets. Notice that no other link function has such
a property.

This mathematical adjustment is simple to explain and to use. But it
has an even more important ramification: if we have multiple models
for different campaigns, then we can argue we can compare their
probability estimates of success and choose accordingly based on their
ex ante, which are more useful, instead of their ex post, which are
too specific. It will become really a prediction and not just a
classification. We will come back to this subject in Section
\ref{sec:overall}.

\section{{\tiny Positive} and {\huge Negative} Set}
\label{sec:positive-and-negative-set}

Events like clicks or any call to action are rare.  If you consider
the training set as the composition of positive events and negative
events, the choice of positives is clearly defined. The choice and the
quantity of negatives is quite a different problem: clicks to other
campaigns, non-clicked impressions but delivered to the same campaign,
to other campaigns, and impressions that are not even selected for
bidding. The list is long and the number of impression in it is very
large. To give a quantitative measure, we may have 1 Million click,
3.5 Billion delivered impressions, and 2 Trillion available
impressions every year. Common practice would be to choose 1 Million
negatives, but {\em where} to find these impressions and {\em what}
are these impressions are very tricky questions: in practice, the
choice of the negative impression to use for training affects any
model.

At first, given a campaign and its clicks, say one thousand, we
thought we could choose another one thousand from the delivered
impressions. Then we could build a model $\tilde{\Vc{\beta}}$. Having
the model, we could score all delivered impressions and create a
distribution. Then we could bin the distribution in 100 bins and
sample 10 impression per bin in order to find a second negative
set. Then compute $\hat{\Vc{\beta}}$.

There is one practical problem: we need to score all impressions and
this is just to retrieve a suitable sample as small as the number of
clicks.

From a practical point of view, we used clicked impressions only:
Given an active campaign from time $t_A$ to $t_B$ with $t_A < t_B$, we
collected all its clicks, these are the positives. The clicks to the
other campaigns in the same interval of time are the negatives. In
practice, we split the click space so that each model and campaign
will bring forth its unique features and all may cover the available
clicks. We shall show the models so created will actually cover this
space.

This is the training set upon we are going to build the model
$\hat{\Vc{\beta}}$. Training and calibration shall be described in
the following section. Once the model is built, we will have a better
understanding what features are important and how they affect the
model. Also, we may have a clear understanding what features determine
a clear rejection: Thus we can estimate a realistic size of available
impressions (ex ante) and thus scale the model accordingly.

\section{Binomial Training and Calibration}
\label{sec:training}

For every campaign, we associate a model. A campaign has positives in
the interval of time $t_A$ to $t_C$ with $t_A < t_C$. We take all
other clicks in the same period of time as negatives. We are bound to
create a training set and a calibration set. We considered two
options: either we split the set by time or by size.

If the campaign has been running for some time and often there are
campaigns running for years, we could choose an instant of time $t_A <
t_B <t_C$ so that $\frac{t_B-t_A}{t_C-t_A} =\frac{3}{4}$. The division
ratio of $\frac{3}{4}$ is arbitrary. The training $T$ is based on the
interval $[t_A,t_B]$ and the calibration $C$ is based on the interval
$(t_B,t_C]$.

However, often campaigns are short spanning a few weeks. The
calibration period would be of only few days of a week and covering
the end of delivery (e.g., less active because we have already reached
our audience or more active because we reach critical mass of
delivery). We could consider the events during the interval $t_A$ and
$t_C$ as a set and choose $T$ and $C$ randomly so that
$\frac{|T|}{|C|}=3$. This latter choice is our default. The training
should have enough information and the calibration should provide an
independent validation.

The former, the distinction between $T$ and $C$ by time, would show
the model predictive capability with the assumption that $T$ is
representative. The latter would show the classification prowess and
the calibration is an independent validation.

\subsection{Receiver Operating Characteristic, ROC.}
\label{sec:roc}

Assume we can build any model using the training set. We measure its
quality using the Calibration set by its ROC. This is a graphical and
quantitative measure. For each element in $\Vc{x}\in C$, compute its
probability of success using the model computed using Equation
\ref{eq:logbetamatrix}: $P[Y_i=1]$ where $Y_i \sim \Vc{x}$. We know
the estimated probability by the model and we know whether or not it
was a true click. Considering $C$, we can compute all probabilities and
sort them from the largest to the smallest: ${\cal P}=\{P[Y_i=1]\}_{\pi(i)}$.

Then, we compute for each $p\in {\cal P}$ the number of true-click
events $Y_i$ such that $P[Y_i=1]\geq p$ over the total number true
clicks: True Positive Rate (TP). Also, we compute the number of
true-no-click events $Y_i$ such that $P[Y_i=1]< p$ over the total
number of no-clicks: False Positive Rate (FP).

For each probability $p$ above, we have two rates $TP(p)$ and $FP(p)$
where $0\leq FP(p),TP(p)\leq 1$. These represent a curve (i.e., the
ROC curve) where the abscissa is $FP$ and the ordinate is $TP$. In
practice, this curve is embedded into a square with unitary side and
left-bottom vertex on the coordinate $(0,0)$ and right-top vertex on
the coordinate $(1,1)$.

If we draw a straight line from $(0,0)$ to $(1,1)$, this is a ROC
curve with a specific meaning: If a model has such a curve, it means
that for every $p$ we have $TP(p)=FP(p)$ and thus we have a constant
probability 1/2 to guess an event right. This looks like a fair coin
flip. In practice, any useful model should provide more information
than a coin toss and its curve must be above this straight line. The
area between these represents the quality of a model. Given a model
$M$ and a calibration set $C$, we represent this area as $ROC(M)_C$.

Thus if we have two models $M_0$ and $M_1$ built from the same
training set $T$ and validate of the same calibration set $C$, we say
that  model $M_0$ is better than $M_1$, when $ROC(M_0)_C>ROC(M_1)_C$.

\section{Binomial Dimensions/Features Exploration} 
\label{sec:dimensions}

In practice, we assume that the explanatory variable-vector $\Vc{x}$
as in Equation \ref{eq:linear} will be able to bring forward the
features necessary to create a binomial model. We need to explore and
choose these explanatory variables and thus quantify their {\em
  explanatory power}. 

So far, Training is used to build the model and Calibration is used to
validate the model.  We have formalized a quantitative measure of
model quality. To explore the feature space, we choose different
spaces and build models, then we compare the models. 

We have available the following feature space: 
\begin{enumerate}
  \item Ad-Exchange: e.g., AppNexus, MoPub.  This represents the set
    of available publishers at our disposal, the different devices and
    different bidding engagements.
  \item Hour of the day: e.g., 17 PM.  Evening hours in the west coast
    have more users than early hours in the east coast. 
  \item Day of the week: e.g., Wednesday. Working days are often more
    engaging than week ends.
  \item Ad format: e.g., video, banner, which represents also the
    location of the ad. Ad Video have more clicks because are more
    difficult to turn off or pause.
  \item Ad size: the real estate size of the ad space. Larger ads
    capture more attention and they are more expensive.
  \item Domains: the sites where the ads are distributed (user
    interests). Most of user targeting is based on the sites we visit
    or particular pages of a site. For example, \verb2yahoo.com2 and
    \verb2finance.yahoo.com2 are two different domains,
    \verb2google.com2 and \verb2google.com/finance2 are not. In
    general, the domain provide a signal (but not always).
  \item Geographical distribution by ZIP: e.g., 95131 (user
    location). There are different levels of precision: IP, lat-long,
    parcel, City, ZIP-4, ZIP, DMA, State. We use ZIP because is an
    intermediary location and it is coarse enough for our purpose. At
    the beginning of this project we considered ZIP and City together,
    we dropped the City because it is a feature harder to compute at
    run time.
 \end{enumerate}

In practice, the first five dimensions describe a limited feature
space: there are only 24 hours in a day. The last two dimensions are
different: there are thousands of ZIPs and there can be thousands of
domains and changing during the year. We need to explore subset of
domains, subset of ZIPs, and we need to understand if there is
correlation. Even if this scenario is specific to our problem, the
properties above cover a wider spectrum of applications.

Before we present what we do, this is what we do not do: we could take
all possible dimensions and cross terms/correlation and train a single
model. If convergence is possible, we could then manually remove
betas: we could start with removing betas associated to rejection,
then removing betas with little contribution. The latter could be
achieved by minimizing the so called $L_1$ (max error) instead of
$L_2$ (variance error) to naturally suppress dimensions. Dimension
suppression will change the overall response of the model and thus the
ROC. We measure the quality of a model by its overall ROC: thus, we
would like to compare their un-altered ROC curves.

Now, the first model we compute is without domains and without ZIPs.

Given the training set, we compute the frequency of domains and ZIPs
for the positive and for the negative cases. For example, we take the
$K{=}10$ most frequent domains that are associated with the positives
and we take the $K{=}10$ most frequent domains associated with the
negatives. Then, we take their union ($K{\leq}20$). We build a model
and we compare with the best built so far using the Calibration
set. We repeat the procedure only for the top $K{=}10$ ZIPs. We repeat
the process with top $K{=}10$ ZIPs and domains. We record the best model
which has the bigger ROC. We do not create a model with $ZIP*Domain$. 

Now, we repeat the process for $K {\in} (20,50, 100, 200)$.  While the
choice of $K$ is arbitrary, the idea is based on the incremental
introduction of more attributes in order to check weather or not they
provide more discriminating information. The exploration is mechanic
and there is no early stop procedure: We do not know if rare features
determine completely the rare positives. However, for computational
and time reason we cannot build a model with all domains and all ZIPs.
We found reasonable to have a maximum of 800 features so that the
model would have no more than one thousand betas. In practice domain
and ZIP are not correlated and rare events have little signal.

In practice, we may have hundreds of campaigns active and we have to
build/rebuild models for them. Every model is used for the
construction of the training and calibration set, however every model
is computed in parallel and independently. It may happen that we
cannot build a model for some campaigns. This failure will not affect
the others models and the other campaigns. In Section \ref{sec:speed},
we will describe the architectural challenges for this process.

\subsection{Threshold Selection.}
\label{sec:threshold}

For each campaign, we have chosen the features based on the quality of
the ROC measure. This is a measure of quality for every probability,
for every case in the calibration set.  If we want to use this model
to choose the impression we would like to bid for, we need to have a
threshold suggesting a specific probability: above the threshold is a
bid, below is a no bid.  We could use the average probability, which
is the CTR of the campaign, but we use a different idea. 

The ROC is the measure of the area above the straight line from (0,0)
to (1,1). The straight line is the representation of the random coin
toss. The model has the least random behavior when it is at its
farthest point from the straight line. This computation is simple and
the meaning is intuitive. 

\subsection{The model.}

In practice, the final model is the composition of: the intercept
$\beta_0$, the set of betas $\{\beta_i\}_{i>1}$ related to the
explanatory features, and the threshold.  Because the function $g()$
is increasing and continuous, we do not need to compute the real
probability.

We accept an impression if $\beta_0+\sum_{i>0}\beta_i >
threshold$. The scoring boils down to a sum of betas, which can be
done quickly. The matching of the impression dimensions with the model
dimensions requires a little more work, but it can always be done by a
binary search (or hashing).

\section{Polytomous Response}
\label{sec:overall}

In this section, we are going to present a few considerations on the
application of these models to real campaigns, impressions, and what
could be the performance at run time for these systems.

\subsection{Coverage.}
\label{sec:coverage}

In isolation, a single model will tend to reduce the number of
acceptable impressions: if we are targeting a rare event, only few
impression will be very likely and the majority will be at best
disputable and rejected. Please, consider that we are modeling events
with an average probability of success of about 0.005, such a model
may choose 8 impressions in 1,000. In practice, one model for one
campaign will choke the delivery. What about 100 models?

We considered a few hundred campaigns deployed in the past and we
modeled about 102 models. Then we took 279,560,699 of delivered
impressions (e.g., one week say). Only 349,585 impressions are
rejected by all models. This means that while each campaign may well
starve, overall they do not. The models re-distribute the impressions
already bought and delivered, 1 model will choke the delivery, 100
will have complete coverage. This means that the current rules used
for the buying and delivery provide the variety and the quantity to
serve all campaigns as a whole.

The models can be used after the decision of bid is taken assuring
that delivery and pacing of campaigns.

\section{Precision and Recall}
\label{sec:precision}

Now, given all clicks can the 102 models recognize them back? In this
section, an impression is taken from the set of clicks, thus we know
what is the campaign associated with any impression.

Let us introduce the following common terminology: given an impression
$x$ and a model $P_{C_j}()$ for campaign $C_j$,

\begin{itemize}
\item A {\em tp} true positive case is when $x$ is a click for
  campaign $C_j$ and $P_{C_j}(x) \sim 1$ (a.k.a. above the threshold).
\item A {\em fp} false positive case is when $x$ is NOT a click for
  $C_j$ and still $P_{C_j}(x) \sim 1$. 
\item A {\em tn} true negative case is when $x$ is NOT a click and
  $P_{C_j}(x) \sim 0$.
\item A {\em fn} false negative case is when $x$ is a click for $C_j$
  and $P_{C_j}(x) \sim 0$.
\end{itemize}
We can then recall the following definition: 

\begin{equation}
  Precision = \frac{\sum tp}{\sum tp+\sum fp}
\end{equation}
\begin{equation}
  NegativeRate = \frac{\sum tn}{\sum tn+\sum fn}
\end{equation}
\begin{equation}
  Recall = \frac{\sum tp}{\sum tp +\sum fn}
\end{equation}
\begin{equation}
  Accuracy= \frac{\sum tp+\sum tn}{\sum tp+\sum fp + \sum tn+\sum fn}
\end{equation}

Having multiple models, it may happen that one impression is vetted
by multiple models. We could give it to the model with the {\em top}
score, or we can provide at random to any of the models with scores
higher than their thresholds, this is like a {\em set} decision. Of
course, the {\em top} and the {\em set} policies count precision and
recall differently: we have different metrics. In Figure
\ref{fig:figure7}, we show a graphical representation of 4 measures
used in the field. The figure is like a time series.

\Singlefigure{1}{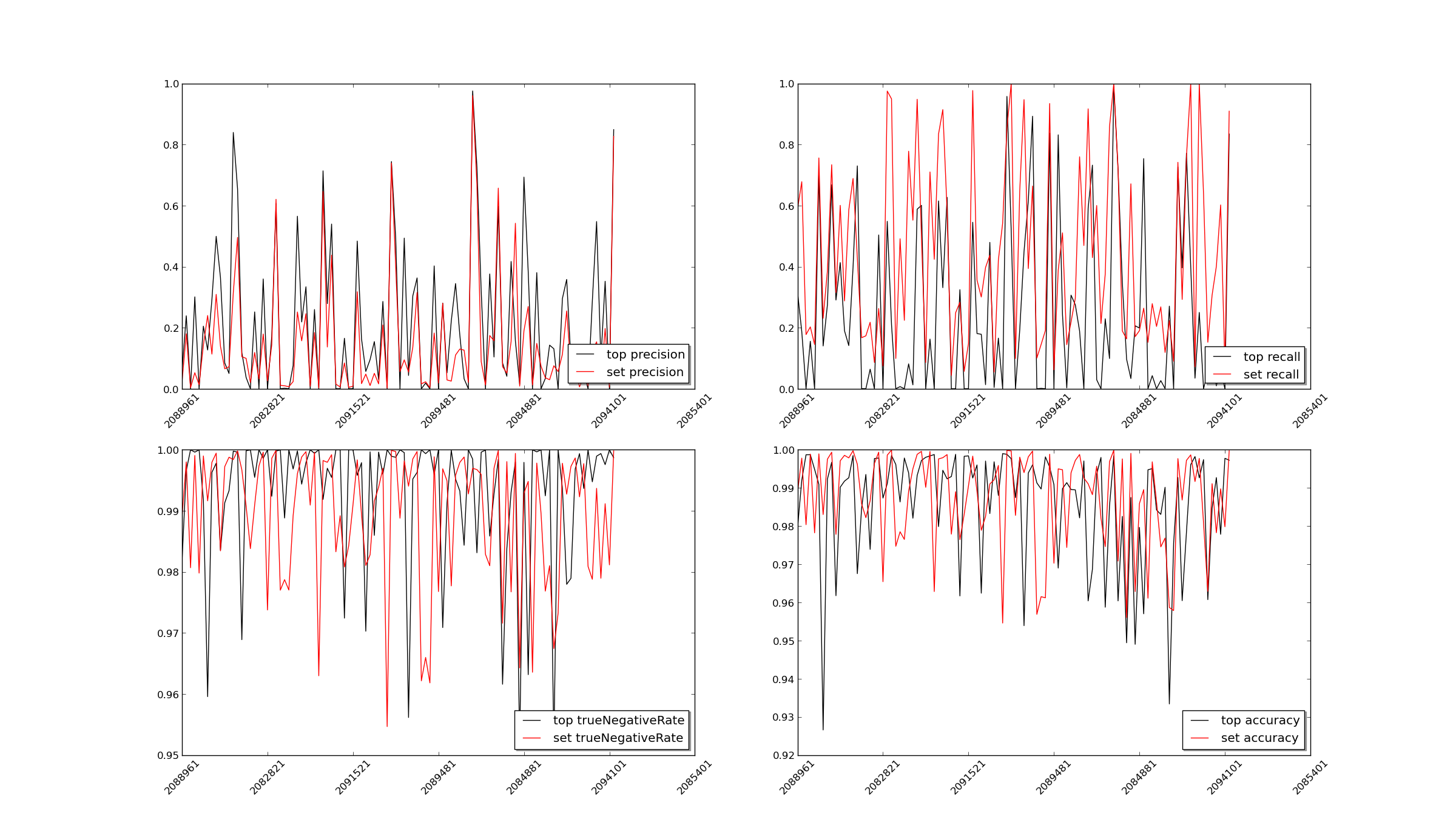}{Precision, Recall, Accuracy and Negative
  Rate}{fig:figure7}

In principle, we could take different thresholds for each models and
determine the configuration maximizing any measure.  In practice and
at run-time, the thresholds will change so that to throttle the
delivery.  This is a hard problem to solve, even to formulate.

Nonetheless, if we now take the definition of true/false
positive/negative, we can consider to compute the Precision and Recall
of the set of models as a single entity: the true positives are the
sum of all models' true positives.

We have: {\em top} Total Precision 0.2945 and Total Recall 0.2917;
{\em Set} Total Precision 0.1300 and Total Recall 0.4103. In practice,
the former has a better precision overall, it can recognize true
positives, but it will increase the false negative. The latter will
have fewer false negative.

We have a measure of the polytomous model, which is composed of binary
models: in practice, every model will have a weight associated to the
importance of the campaign (e.g., money or total number of impression
to deliver). 

Clearly the building of each model separately is appealing because we
can turn them off without need to retraining the others.

\subsection{Scoring Speed.}
\label{sec:speed}

The scoring in itself is the sum of betas. The sum can be done very
quickly as soon as we know which betas to use. Given an impression
$x$, each model betas are different and some beta may not have the
betas associated with $x$'s features.

How many impression we can score per second or QPS (query per second)?

We implemented a multithreaded library written in C for the scoring
above. We chose C because we wanted to measure performance in two
systems: Intel Phi coprocessor with 50 cores (with four thread each
for a total of 200 cores) and 2 xeon Westemere processors each with 6
cores with two thread each for a total of 24 core.

A single core on the Westemere can provide 10,000 QPS (having all
impressions in memory already). Both systems can provide steady 1
Million QPS making the scoring affordable at run time; note, we moved
all impressions to the internal memory in the Phi system before
measuring the throughput. Using in combination, we can achieve twice
as much. This test is designed to compute the peak throughput and not
the minimum or maximum latency, which is more common for real time
bidding.  Considering the highly parallel Phi that can achieve 3
TFLOPS sustained performance, with 300Watt consumption, 8 GB of memory
and much more cache memory to assist the internal cores, we must admit
that the server configuration with 2 xeon processors is a better
choice (160 Watts and 64 GB memory). This is because the scoring
function uses very little the deep pipeline avail in the Phi, which is
the real reason of its high peak performance.

Unfortunately, we could not test the performance on GPUs such as
Radeon 290 (available in the same system) because of time constraints
and unavailable resource to export the library as OpenCL kernels.

\section{Conclusions}
\label{sec:conclusions}

Training polytomous models composed of binary models is appealing for
campaign optimization. In this work, we show how to scale to hundreds
of binary models (hundreds on independent responses). This is to show
the applicability of the theory developed twenty years ago. 

In practice, multiple campaigns can be optimized at the same time. Our
ability to deploy multiple models circumvent a few critical issues
about binary models and we can measure the quality of each campaign in
isolation and as a collective. The collective set of models and each
model can be modified at any time without affecting the others or the
scoring speed. 

Each model is built on top of the unique features that describe the
campaign among the other campaigns.

\section{Acknowledgments} 
\label{sec:ack}
The work presented here was mostly done while the author was at
Brand.net.

We would like to thank several bright(er than us) people: Aram Campau,
David Folk, Christofer Gilliard, Konstantin Bay, James Tsiao, and Yang
Li. They helped starting and they nurtured this project. The
inspiration to write our contributions stems from the work by Lee et
al.  \cite{LeeODW2012} and their application in their real time bid
optimizations \cite{LeeJD2013}.

\bibliographystyle{abbrv}
\bibliography{pa.strass2}

\end{document}